\documentclass[a4paper,twocolumn,aps,prx]{revtex4-2}
\usepackage{graphicx}% Include figure files
\usepackage{dcolumn}% Align table columns on decimal point
\usepackage{bm}
\usepackage{color}
\usepackage{amssymb}
\usepackage{amsmath}
\usepackage{amsmath}
\usepackage{xfrac}
\usepackage{mathrsfs}
\usepackage{comment}
\usepackage{titlesec}
\usepackage{lipsum}
\usepackage{graphicx}
\usepackage{subcaption}
\usepackage{float} % optional (for H placement)
\begin{document}

\preprint{APS/123-QED}

\title{Extreme Events in an Active Fluid Medium}
\author{Joydeep Das}\email{mp21022@iisermohali.ac.in}
%  \textbackslash\textbackslash
\author{Abhishek Chaudhuri}\email{abhishek@iisermohali.ac.in}
\author{Sudeshna Sinha}\email{sudeshna@iisermohali.ac.in}
\affiliation{%
Department of Physical Sciences, Indian Institute of Science Education and Research Mohali\\ Sector 81, SAS Nagar, Punjab 140306, India
%  \textbackslash\textbackslash
}%% \altaffiliation[Also at ]{Physics Department, XYZ University.}%Lines break automatically or can be forced with \\
%% \email{Second.Author@institution.edu}
%\affiliation{%
% Authors' institution and address\\
%  \textbackslash\textbackslash
%\affiliation{
%\affiliation{


\date{\today}% It is always \today, today,

\begin{abstract}
We observe the emergence of extreme events in an active fluid involving two distinct chemical species that regulate active stress. One species is slow diffusing and the other is fast diffusing, and the growth of the fast-diffusing species is modelled using a nonlinear logistic term. We demonstrate the presence of extreme events in the temporal evolution of the concentrations, as well as in the spatial profile of the system, 
%through examination of the time-series, bifurcation diagrams and the probability distribution functions of the concentration of the two species,
%. We explore the change in extreme event occurrences under varying P\'eclet number and strength of the nonlinear growth term, and find that the rare and uncorrelated extreme concentration build-ups are found 
in regimes of merging-emerging soliton-like dynamics and spatio-temporal chaos, through analysis of the time-series, bifurcation diagrams,  probability distribution functions of the concentration of the two species, return maps and distributions of inter-event intervals. Interestingly, we also find evidence of pronounced bunching of extreme events and super-extreme events in the slow chemical species in the soliton-like regime. We go on to systematically explore the dependence of the extreme event occurrences on the P\'eclet number and the strength of the nonlinear growth term, and find that the probability of extreme events increases after a critical P\'eclet number, while increasing the nonlinearity suppresses extreme events. Lastly, in order to gain further insight, we investigate a modified mode-truncated reduced order model comprising of coupled differential equations mimicking this active fluid system. We find that this reduced order model also exhibits extreme events whose emergence is correlated with a sudden expansion in attractor size due to a crisis arising from attractor collision. 
So these results demonstrate the existence of extreme events in an active fluid system, and are of potential relevance to biological phenomena where active transport plays an important role.
%and provides insight into their emergence through a reduced order model.
%Thus we demonstrate the existence of spatial and temporal extreme events in an active fluid system, indicating the ubiquity of emergent extreme events in complex systems.
\end{abstract} 
%\keywords{Suggested keywords}%Use showkeys class option if keywordhttps:
\maketitle

%\tableofcontents
\section{Introduction}

%Pattern formation is an integral part of the development of biological systems. 
The classical framework for understanding pattern formation in natural systems is that of reaction-diffusion models \cite{turing1990chemical,cross2009pattern,cross1993pattern,kondo2010reaction,green2015positional}, 
%The emergent patterns observed in this broad class of systems include time-independent and time-dependent oscillatory patterns~\cite{turing1990chemical}, symmetry-breaking instabilities~\cite{prigogine1968symmetry}, travelling waves~\cite{zaikin1970concentration}, spirals~\cite{winfree1972spiral} and jumping oscillations~\cite{knobloch2021origin}. Importantly, 
and many of the patterns observed in such model systems have also been found in experiments~\cite{ouyang1991transition,castets1990experimental}. 
Now, active transport plays a crucical role in pattern formation, particularly in biological systems. So an important approach to modelling biological systems integrates the contributions of both mechanical and as well as chemical effects~\cite{harris1984generation}, for instance 
%actin networks showing non-equilibrium dynamics by force generation through myosin motor activity~\cite{kumar2014pulsatory,juelicher2007active,ramaswamy2010mechanics} and pattern formation in 
an active fluid medium in the presence of diffusing chemical species that are advected by self-generated flows produced by concentration-dependent active stress gradients \cite{bois2011pattern,kumar2014pulsatory}. 
%Cell crawling is also dependent on the density of myosin motors, and so it can be studied by considering the distribution of myosin motors as a supercritical van der Waals (vdW) fluid~\cite{drozdowski2023optogenetic}. Spontaneous protrusion dynamics is modelled by mechano-chemical coupling via a polymerizing active gel layer~\cite{levernier2020spontaneous,laplaud2021pinching}.

Specifically, the actomyosin cortex, which lies just beneath the cell membrane, consists of actin filaments and myosin motor proteins which crosslink the actin. This generates mechanical forces, giving rise to an active stress component in a thin mesoscopic layer of the actomyosin cortex. The actomyosin cortex is treated as an active fluid at time scales of morphogenesis, which is an essential part of organismal development, relevant to human tissue development as well as development in systems such as worms, flies, zebrafish and mice~\cite{kruse2024actomyosin}. The active stress can now be considered to be a function of the concentrations of regulatory chemical species to complete the mechanochemical integration, leading to pattern formation even in the absence of chemical reactions~\cite{bois2011pattern}. When extended to two chemical species which undergo advection-diffusion and regulate the active stress, this framework gives rise to pulsatory patterns~\cite{kumar2014pulsatory}. Further, spontaneously emerging localized states in the active fluid medium have led to the understanding of localized cellular patterns~\cite{barberi2023localized},
%. It has been shown that localized states can emerge spontaneously, 
analogous to isolated clusters of actin and signalling molecules in cancer cells~\cite{dd2061aa35da449e91a2a6209d0118ab}.
%when the assembly of active matter is regulated by the presence of chemical species that are advected with flows resulting from active stress gradients. 

In another research direction, the study of extreme events has gained momentum in recent years. Initially studied in the context of oceanography as rogue waves~\cite{pelinovsky2008extreme} and climatology~\cite{ross2003climatology} as abrupt shifts, the concept of extreme events has expanded to fields including optics, electronics and mechanical oscillators~\cite{kingston2017extreme,pisarchik,thangavel2021extreme,chen2020predicting,spitz2020extreme}, as well as neuronal models~\cite{roy2023impact,roy2024impact}. Extreme events are very relevant, due to their capacity to inflict serious destruction, such as in  earthquakes~\cite{pisarenko2003characterization}, floods~\cite{buchele2006flood}, droughts~\cite{hoerling2013anatomy}, epidemic spreading~\cite{mcmichael2015extreme} and solar flares~\cite{buzulukova2017extreme}. These events are recognized as significantly large-amplitude deviations from long-term nominal behavior. The large fluctuations are short-lived but appear in a recurrent manner~\cite{das2024complexity,chowdhury2022extreme}, and have magnitudes that are several times larger than the standard deviation of the data set~\cite{sudharsan2025extreme1,ray2022extreme,ray2020extreme,ansmann2013extreme,shashangan2025propagation1,hariharan2026heterogeneous1,thangavel2026emergence,hariharan2026synchronization,ray2019intermittent,kanagaraj2026extreme,knobloch2024emergence}.
%Lots of studies are already being performed on extreme events but Extreme events are not limited only to these topics. 

In this current study, we explore extreme events in an active fluid medium in the presence of two distinct chemical species. One of the species diffuses fast and the other one diffuses slowly, and the fast-diffusing species experiences a nonlinear logistic growth. In an earlier work, it was shown that this system can exhibit diverse of patterns under the variation of parameters. These include stationary patterns, pulsatory patterns, symmetry breaking chimera states and merging-emerging dynamics~\cite{das2026order}. In the sections below, we first describe the model, and then we demonstrate the existence of extreme events in the temporal evolution of the concentrations, as well as in the spatial profile of the system, through examination of the time-series, bifurcation diagrams and the probability distribution functions of the concentration of the two species, in regimes of merging-emerging soliton-like dynamics and spatio-temporal chaos. We also find clear evidence of bunching of extreme events, through investigations of the inter-event interval distributions and return maps. We explore the change in extreme event occurrences under varying P\'eclet number and strength of the nonlinear growth term, and find that the probability of extreme events increases after a critical P\'eclet number, while increasing the nonlinearity suppresses extreme events. Finally, in order to gain further understanding of these extreme events, we investigate a mode-truncated reduced order model of the system, and demonstrate that it also exhibits temporal extreme events. 
%In our system, we observe localized soliton like structures, those are moving and merging-growing in an irregular fashion. Here, we have found some instances in space and time where they exceed the nominal behavior and exhibit extreme event dynamics. In this work first we will discuss about the mathematical model of the system, then we discuss briefly the linear stability analysis of the system, after that we discuss on the probability of extreme events , inter-event intervals and lastly we discuss a mode truncated version of this system, that can also exhibit the extreme events. We hope that our work will give a direction to the exploration of extreme events in active matter systems. 

\section{Model of the Active Fluid system}

Consider two distinct chemical species ($A, I$) regulating the active stress in a one-dimensional active fluid system in a thin-film geometry. The important quantities in the system are the concentration fields of $A(x,t)$ and $I(x,t)$ at time $t$ and position $x$. The evolution equations of the two chemical species in one-dimension are given as:
%The new significant feature we will incorporate in our model is a logistic growth for the activator chemical species. So the activator and inhibitor concentrations evolve according to the advection-diffusion equations, through a spatial gradient operator, with an additional logistic growth term, as described below: 
\begin{eqnarray}
\partial_t A &=& - \partial_x (vA) + D\partial^2_x A + rA(1 - A/K) \nonumber \\ 
\partial_t I &=& - \partial_x (vI) + \alpha D\partial^2_x I
%\frac{\partial A}{\partial t} = -\frac{\partial (vA)}{\partial x} + D\frac{\partial^2 A}{\partial x^2} + r A (1-\frac{A}{K})
\end{eqnarray}
%\begin{equation}
%\frac{\partial I}{\partial t} = -\frac{\partial (vI)}{\partial x} + \alpha D\frac{\partial^2 I}{\partial x^2}.
%\end{equation}
Thus, both species exhibit advection and diffusion, with the diffusive component determined by the diffusion coefficient $D$ and the advective component given by the bulk fluid flow velocity $v$. The relative diffusion coefficient of the two species is determined by the parameter $\alpha > 0$.
%is the ratio of the diffusion coefficients of the the activator and inhibitor species, and it is considered to be small here, i.e. we consider a fast diffusing activator and a slow diffusing inhibitor. 
The nonlinear logistic growth term, with strength $r$, is a generic reaction term that has the capacity to destabilize a steady state with zero concentration and also saturates at a finite carrying capacity $K$. For the active fluid at low Reynolds numbers, the inertial terms can be neglected, and the force balance equation gives
\begin{equation}
\partial_x\sigma = \gamma v,
%\frac{\partial \sigma}{\partial x} =\gamma v.
\end{equation}
with
\begin{equation*}
\sigma = \eta\partial_x v + \sigma_a
%\sigma = \eta\frac{\partial v}{\partial x} + \sigma_a
\end{equation*}
\\
giving the total stress. $\sigma$ consists of a viscous stress component $\eta\partial_x v$ with $\eta$ being the viscosity of the medium and an active stress component $\sigma_a$ regulated by the concentrations of the chemical species:
\begin{equation*}
\sigma_a = \sigma_0 f(A,I)
\end{equation*}
where $\sigma_0$ is an active stress amplitude, and $f(A, I)$ is a dimensionless function describing the regulations of the active stress:
%As proposed in Ref.~\cite{kumar2014pulsatory}, we choose:
\begin{equation}
f(A,I)= (1+\beta)\frac{A}{A+A_S}+(1-\beta)\frac{I}{I+I_S}
\label{f_A_I}
\end{equation}
where $\beta$ is an asymmetry parameter, and $A_S$ and $I_S$ represent the saturation values of the concentrations of the two chemical species~\cite{kumar2014pulsatory}. For $\beta < -1$, $A$ down-regulate and $I$ up-regulates active stress; for $-1\leq \beta \leq 1$ both species up-regulates stress and for $\beta > 1$, $A$ up-regulates while $I$ down-regulates active stress.

The equations above can be expressed in dimensionless form as:
\begin{eqnarray}
    \partial_t A&=&- \partial_x (vA) + \partial^2_x A + RA(1 - A) \nonumber \\ 
\partial_t I&=&- \partial_x (vI) + \alpha \partial^2_x I \nonumber \\
\partial^2_x v &=&v - Pe \ \partial_x f(A,I)
\end{eqnarray}
where $A, I, x, t$ and $v$ are now non-dimensional. The two important parameters in the system are as follows: (i) scaled non-linear growth parameter $R = r\eta/\gamma D$, which reflects the strength of the nonlinear growth term, and (ii) dimensionless P\'eclet number $Pe = \sigma_0/\gamma D$, which reflects the ratio of the advective transport to diffusive transport in fluid flow.
%In this work we consider $(A_S,I_S) = (3A_0,3I_0)$, where $(A_0,I_0)$ are the homogeneous steady state concentrations.

This system was examined in Ref.~\cite{das2026order}, through extensive numerical simulations, as well as linear stability analysis. The study showed that this active fluid system exhibited a diverse array of patterns. The salient features were as follows: Increasing P\'eclet number destabilised the uniform steady state. On the other hand, an increase in the nonlinear growth parameter of $A$ expanded the homogeneous steady-state regime. Asymmetry between the species also influenced the dynamics, with low asymmetry failing to produce oscillatory instability. In the parameter regimes of instability, the system yielded patterns ranging from irregular, arrhythmic dynamics at high P\'eclet numbers, to symmetry-breaking chimera states. Further, soliton-like structures where aggregations of species $A$ merge, and new aggregations spontaneously emerge, were also found. So this system offers a rich test-bed of dynamical spatiotemporal patterns for further exploration, where one can potentially unearth other interesting phenomena.

%\begin{figure}
    %\centering
    %\includegraphics[width=0.8\linewidth]{PDEbifurcationsb3R1a0.1Taken700-800aroundx=70.png}
    %\caption{\raggedright Bifurcation diagram displaying the local maxima (blue) and local minima (red) of $A$ at a representative spatial location, with respect to P\'eclet number $Pe$, with $\beta=3,R=1,\alpha=0.1$.
    %at $x=70$ for $T=700-800$ , blue dots represent local maxima and red dots represent the local minima. It clearly shows 
    %The transition from oscillatory state to chaotic state is evident. Further, large deviations from the mean are seen to occur around $Pe \sim 43-48$}
    %\label{bifurcation}
%\end{figure}

\begin{figure*}[htbp]
\centering

% ---------- Row 1 ----------
\makebox[\textwidth][c]{%
\includegraphics[width=0.22\textwidth]{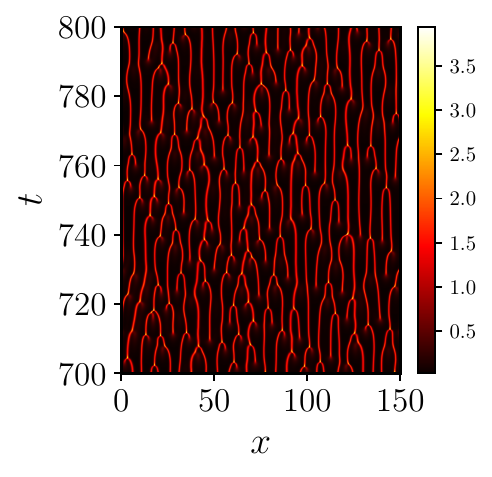}
\includegraphics[width=0.22\textwidth]{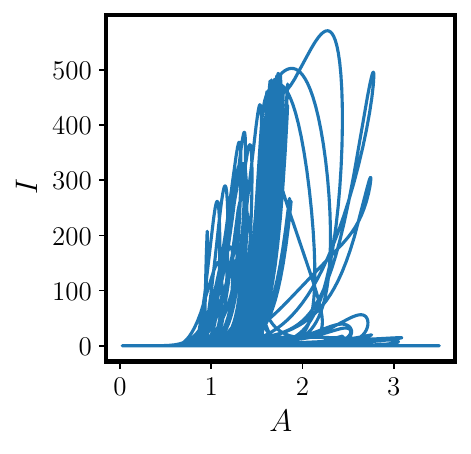}
\includegraphics[width=0.22\textwidth]{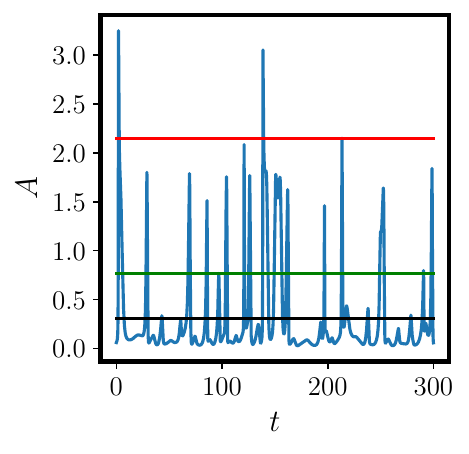}
\includegraphics[width=0.22\textwidth]{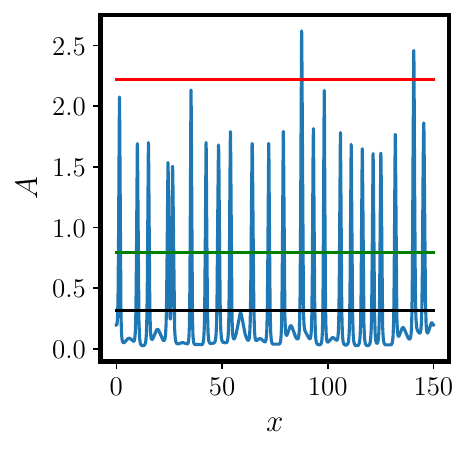}
}

\vspace{0.2cm}

% ---------- Row 2 ----------
\makebox[\textwidth][c]{%
\includegraphics[width=0.22\textwidth]{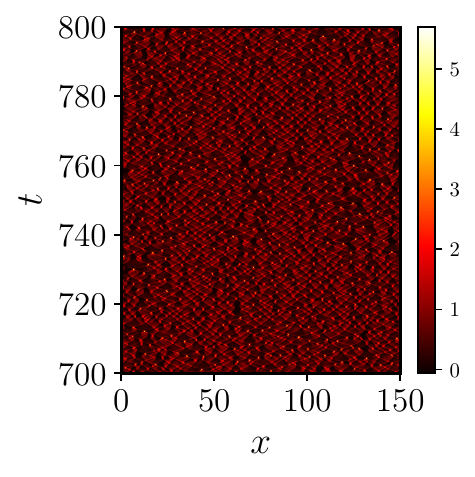}
\includegraphics[width=0.22\textwidth]{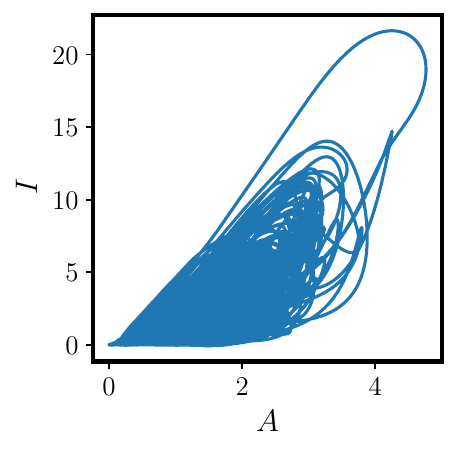}
\includegraphics[width=0.22\textwidth]{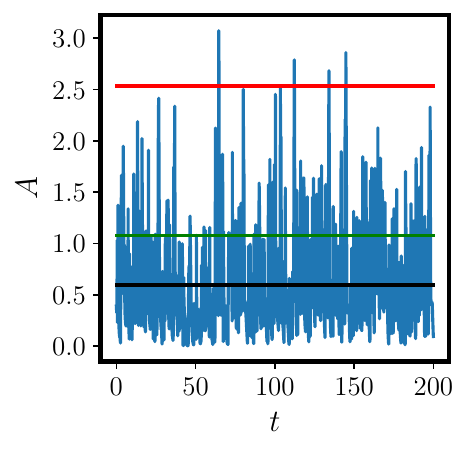}
\includegraphics[width=0.225\textwidth]{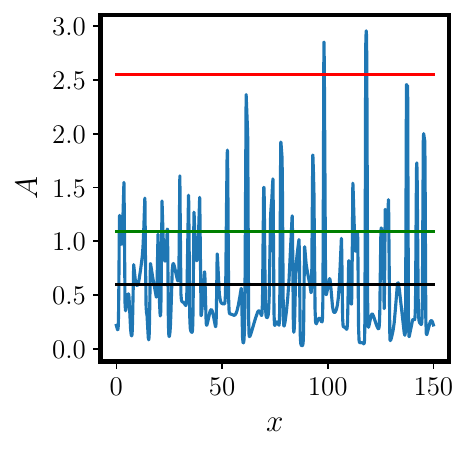}
}

\caption{\raggedright (left to right) Kymographs, phase portraits, time series and spatial snapshots of species $A$, for two representative dynamical regimes: (top panels) merging-emerging soliton-like dynamics, arising for parameters $Pe=24$, $\beta=1.2$, $R=1$, $\alpha=0.1$, and (bottom panels) spatio-temporal chaos, arising for parameters $Pe=48$, $\beta=3$, $R=1$, $\alpha=0.1$. Here the average value $\mu$ of $A$ is shown with a black line, and $\mu+\sigma$ and $\mu + 4\sigma$ (where $\sigma$ is the standard deviation) with green and red lines respectively. }
\label{kymo}
\end{figure*}

\section{Emergence of extreme events}

We probe the emergent spatiotemporal patterns in this active fluid system, through extensive numerical simulations, over a range of parameters, searching for the existence of regimes yielding extreme events. Specifically, we use the pseudo-spectral numerical scheme to evolve the system. With no loss of generality in the qualitative dynamics, we use $2048$ and $4096$ spatial data points for a system of length $L=150$, and time step $\Delta t=0.001$. The saturation values of the concentrations of the two chemical species in Eqn.~\ref{f_A_I} is taken to be $A_S=I_S=3$ \cite{kumar2014pulsatory}. We initialize the system to the homogeneous steady state $(A,I) = (1,1)$, and then investigate the spatiotemporal dynamics of the system following small uniformly distributed random perturbations. 

%In particular, such a dynamical regime occurs for the parameter set: $Pe=48$, $\beta=3$, $R=1$ and $\alpha=0.1$. 
The key  finding of our work here is that we detect temporal and spatial extreme events in the dynamical regime of spatiotemporal chaos and in regimes where soliton-like waves emerge, move and merge, appearing to pass through each other in an irregular manner. The extreme events manifest as rare and uncorrelated extreme concentration build-ups, as clearly evident in the kymographs, time series data, spatial profiles and phase space attractors shown in Fig.~\ref{kymo}, for representative sets of parameters. 

The central observations are as follows: When we examine the time evolution of the concentrations of the two species $A$ and $I$, at a generic site in the system, over long time, we find that these concentrations exhibit sudden large deviations from their mean values at certain uncorrelated instances of time. These huge excursions in the state of the system are rare, but recurrent. So such occurrences can clearly be considered as extreme events in time arising in the evolution of the concentrations of the species. Additionally, when we examine the spatial profile at an instant of time, we find that there are a few uncorrelated sites where the values of the concentrations $A$ and $I$ spike very significantly beyond the average concentration across the spatial profile. These indicate the presence of extreme events in space. These observations are also supported by the geometry of the phase space attractors of the system, which exhibit occasional occurrences of long excursions in phase space, significantly far from the dense region of the attractor where the system is confined on an average. Specifically, in this work, we prescribe the threshold $H$ for the detection of an extreme event to be $\mu+4\sigma$, where $\mu$ is the average and $\sigma$ is the standard deviation. This threshold is marked in the figures, as a reference.

\begin{figure*}
    \centering
    \includegraphics[width=\linewidth]{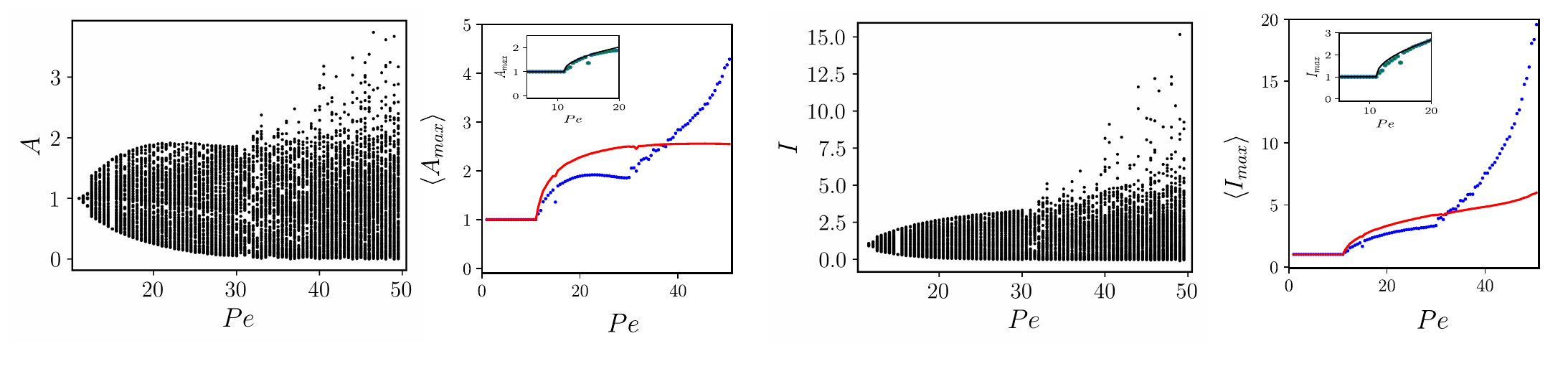}
    \caption{\raggedright Bifurcation diagram displaying the local maxima and local minima of chemical species $A$ (left) and $I$ (right) at a representative spatial location, with respect to P\'eclet number $Pe$, with $\beta=3$, $R=1$, $\alpha=0.1$. 
    %Here red dots represent the threshold values at this site for each $Pe$. 
    Alongside we show the site averaged value of global maxima $\langle A_{max} \rangle$ and $\langle I_{max} \rangle$ of the chemical species with respect to $Pe$ (blue dots), as well as the threshold $H = \mu + 4 \sigma$ for determining extreme events (red curve). In the inset we show the best fit (black curve) of $\langle A_{max} \rangle$ to the functional form 
    %$A_{h}=A_0 + c_1
    $\sqrt{Pe-Pe^c}$
    %and $I_{h}=I_0 + c_2\sqrt{Pe-Pe_c}$, 
    near the bifurcation point $Pe^c$.}
    %Here, $c_1=0.58,c_2=0.3,Pe_c=11$.}
    \label{Bifurcationwithscaling}
\end{figure*}

In Fig.~\ref{Bifurcationwithscaling} we show bifurcation diagrams displaying the local maxima and local minima of chemical species $A$ and $I$ at a representative spatial location, with respect to P\'eclet number $Pe$. Alongside the bifurcation diagrams we also show the site-averaged global maxima $A_{max}$ and $I_{max}$ of the chemical species with respect to $Pe$, as well as the threshold $H = \mu + 4 \sigma$ for determining extreme events. The P\'eclet number at which this global maxima crosses the threshold $H$ marks the point of inception of extreme events. Interestingly, the critical $Pe$ at which the extreme events commence is different for the two chemical species. In the inset we show the best fit of $A_{max}$ and $I_{max}$ to the form $\sqrt{Pe-Pe^c}$,  near the bifurcation point $Pe^c$, consistent with the scaling of oscillation amplitude after a supercritical Hopf bifurcation~\cite{das2026order}. The threshold $H$ also scales approximately as a square root of the difference of the P\'eclet number from the critical value.

Two trends are noteworthy: First, notice that the threshold level $H$ is a smooth curve which rises slowly with P\'eclet number throughout the parameter range. However, the global maxima of the two species display an abrupt change, after which their values climb significantly faster than that of the threshold. This rapid increase in $A_{max}$ and $I_{max}$ can be fit to a third-order polynomial, and so it is markedly different from the square root scaling closer to the supercrictical Hopf bifurcation. The abrupt switch from slowly varying square-root scaling to rapid increase of the global maxima of the species concentrations occurs at the attractor expansion point, as evident through comparisons with the bifurcation diagram. This important difference in the profile of the threshold and the global maxima gives rise to extreme events, with extreme events emerging after the $A_{max}$/$I_{max}$ curves cross the threshold curve. Also note that the rarity of extreme events is reflected in the fact that the scaling of the growth of the threshold is similar throughout the parameter regime, as the very large excursions do not significantly affect this quantity since they are very infrequent and constitute a very low density of points on the attractor. Therefore the extreme events have very low contributions to the mean and standard deviation estimated over long times, as those quantities are dominated by the high density regions of the attractor which contains almost all the phase points.
    
    %Here, $c_1=0.43,c_2=0.74$.

%\begin{figure}
   % \centering
   % \includegraphics[width=0.67\linewidth,height=4cm]{AmaxandAminvsPeforbeta3randomICSaround(1,1)seed42fixedx103P10to50with0.5interval.png}
%    \caption{\raggedright Bifurcation diagram displaying the local maxima (blue) and local minima (red) of $A$ at a representative spatial location, with respect to P\'eclet number $Pe$, with $\beta=3,R=1,\alpha=0.1$.
%    %at $x=70$ for $T=700-800$ , blue dots represent local maxima and red dots represent the local minima. It clearly shows 
%    The transition from oscillatory state to chaotic state is evident. Further, large deviations from the mean are seen to occur around $Pe \sim 43-48$}
%    \label{bifurcation}
%\end{figure}
%\begin{figure}
%    \centering
    %\includegraphics[width=0.7\linewidth,height=4cm]{ImaxandIminvsPeforbeta3randomICSaround(1,1)seed42fixedx103P10to50with0.5interval.png}
   % \caption{\raggedright Bifurcation diagram displaying the local maxima and local minima (black) of chemical species $A$ (top) and $I$ (bottom) at a representative spatial location $x=103$, with respect to P\'eclet number $Pe$, with $\beta=3,R=1,\alpha=0.1$. Here, red dots represent the threshold values at this site for each $Pe$.
    %at $x=70$ for $T=700-800$ , blue dots represent local maxima and red dots represent the local minima. It clearly shows 
    %The transition from oscillatory state to chaotic state is evident, after which large deviations from the mean are seen to occur.
    %around $Pe \sim 43-48$.
   % }
   % \label{bifurcation2}
%\end{figure}

\bigskip
\section{Statistical Properties of Extreme Events}

\begin{figure}[htbp]
%\centering

% ---------- Row 1 ----------
%\makebox[\textwidth][c]{%
%\includegraphics[width=0.22\textwidth]{P(A)vsAforP24b1_2for5crorepointsuptoT50000seed42.pdf}
\includegraphics[width=0.22\textwidth]{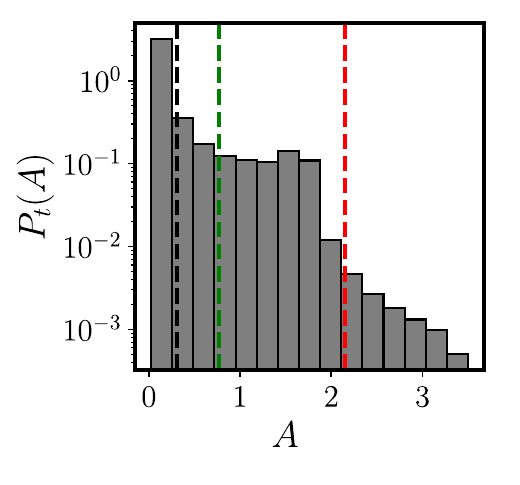}
\includegraphics[width=0.22\textwidth]{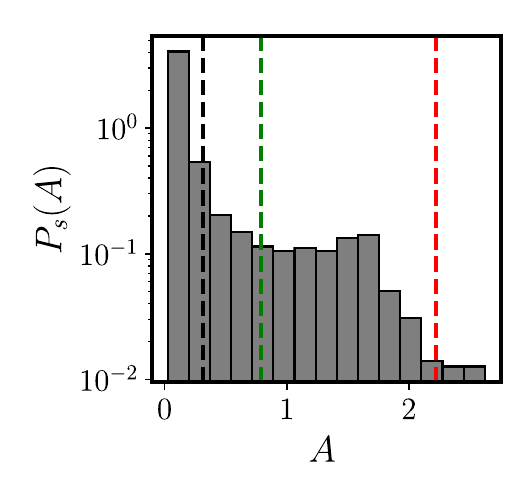}
%\includegraphics[width=0.22\textwidth]{P(I)vsIP48b3R1atx70uptoT5000.pdf}
%\includegraphics[width=0.22\textwidth]{P(I)vsIP48b3R1atx70uptoT5000spatialEE.pdf}
%}

\vspace{0.2cm}

% ---------- Row 2 ----------
%\makebox[\textwidth][c]{%
%\includegraphics[width=0.22\textwidth]{P(I)vsIforP24b1_2for5crorepointsuptoT50000seed42fixed.pdf}
\includegraphics[width=0.22\textwidth]{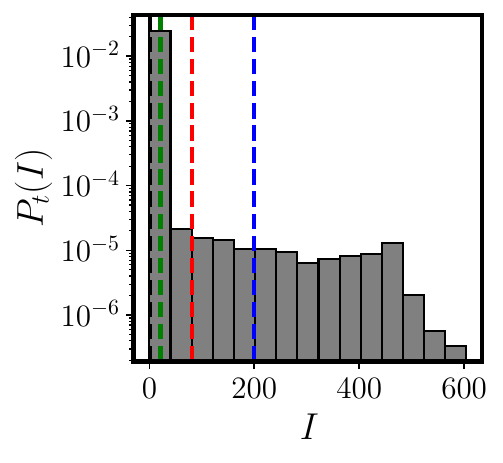}
\includegraphics[width=0.22\textwidth]{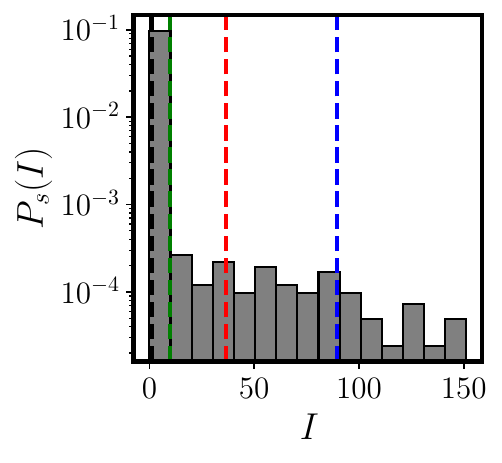}
%\includegraphics[width=0.215\textwidth]{AvstP48b3R1atx70uptoT5000.pdf}
%\includegraphics[width=0.225\textwidth]{AvsxP48b3R1att480uptoT500.pdf}
%}

\caption{\raggedright Probability distribution of the values of the variables $A$ (top) and $I$ (bottom): (left) tracked 
%around 5 crores time steps 
over a long period of time ($\sim 10^4$) after initial transience,
%(removing 50000 steps)
at  a specific spatial location, denoted by $P_t (A)$ and $P_t (I)$ (right) tracked over space, at a specific time instant, denoted by $P_s (A)$ and $P_s (I)$. Here
%$T=5000,L=150$.Here,
%Left side represents the probability distribution of $A,I$ at $x=70$ and Right side represents the probability distribution of $A,I$ at $t=480$. Here, $N_x=4096,T=50000,L=150$(Ignoring the transients).Here, 
$Pe=24$, $\beta=1.2$, $R=1$ and $\alpha=0.1$.
%and $N=4096$. 
The average value $\mu$ is shown with a black line, and $\mu+\sigma$ and $\mu + 4\sigma$ (where $\sigma$ is the standard deviation) with a green and red line respectively. The $\mu + 10 \sigma$ threshold is shown in blue (if this threshold lies within the range displayed in the panels). It is clearly visible that both probability distributions of $I$ extend beyond the $\mu + 10 \sigma$ threshold, indicating the occurrence of super-extreme events.
Here $4096$ spatial data points have been taken for a system of length $L = 150$.
Qualitatively similar features are obtained for $2048$ spatial data points as well.} %Here we have taken random initial perturbations w.r.t the $(1,1)$ state.}
\label{prob_1}
\end{figure}

\begin{figure}[htbp]
%\centering

% ---------- Row 1 ----------
%\makebox[\textwidth][c]{%
%\includegraphics[width=0.22\textwidth]{P(A)vsAforP48b3for50lakhpointsuptoT50000seed42fixed..pdf}
%\includegraphics[width=0.22\textwidth]{P48b3N2048AtemporalEEseed42x103withsymbol.pdf}
%\includegraphics[width=0.23\textwidth]{P(A)_tvsAforP48b3seed42atx103T50to5000N2048.pdf}
\includegraphics[width=0.23\textwidth]{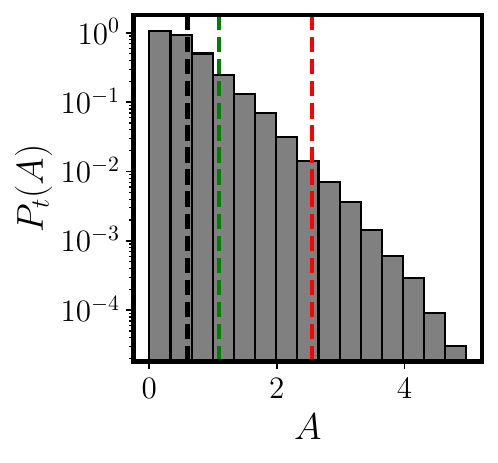}
\includegraphics[width=0.23\textwidth]{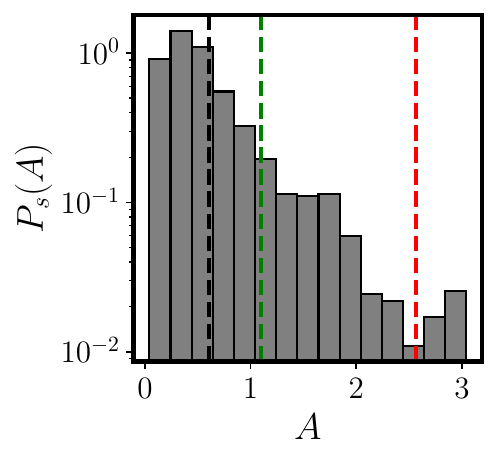}
%\includegraphics[width=0.22\textwidth]{P(I)vsIP48b3R1atx70uptoT5000.pdf}
%\includegraphics[width=0.22\textwidth]{P(I)vsIP48b3R1atx70uptoT5000spatialEE.pdf}
%}

\vspace{0.2cm}

% ---------- Row 2 ----------
%\makebox[\textwidth][c]{%
%\includegraphics[width=0.22\textwidth]{P(I)vsIP48b3R1atx103uptoT5000seed42fixed.pdf}
%\includegraphics[width=0.22\textwidth]{P48b3N2048Ipdfwithsuperextremethresholdx103seed42withsymbol.pdf}
\includegraphics[width=0.23\textwidth]{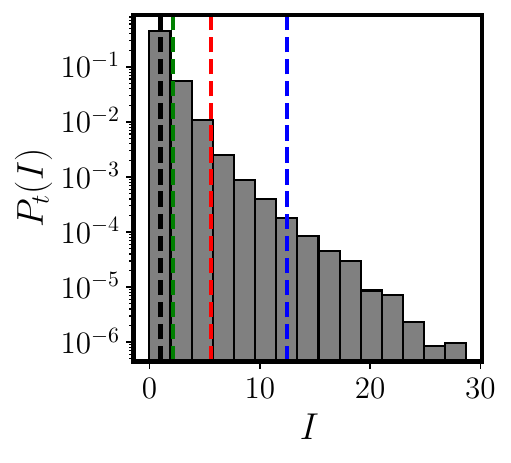}
\includegraphics[width=0.23\textwidth]{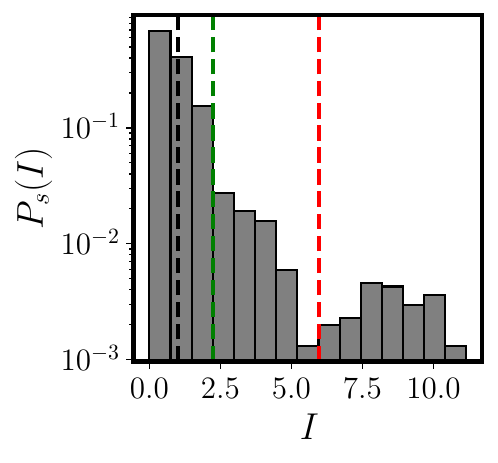}
%\includegraphics[width=0.22\textwidth]{P(I)vsIatt450P48b3N2048seed42fixedspatial.pdf}
%\includegraphics[width=0.215\textwidth]{AvstP48b3R1atx70uptoT5000.pdf}
%\includegraphics[width=0.225\textwidth]{AvsxP48b3R1att480uptoT500.pdf}
%}

\caption{\raggedright Probability distribution of the values of the variables $A$ (top) and $I$ (bottom): (left) tracked over 
%5000 time steps 
a long period of time ($\sim 10^4$) after initial transience, at  a specific spatial location, denoted by $P_t (A)$ and $P_t (I)$; (right) tracked over space, at a specific time instant, denoted by $P_s (A)$ and $P_s (I)$. Here  
%$T=5000,L=150$.Here,
$Pe=48$, $\beta=3$, $R=1$ and $\alpha=0.1$.
%, and $L=150$.
%and $N=2048$. Here t
The average value $\mu$ is shown with a black line, and $\mu+\sigma$ and $\mu + 4\sigma$ (where $\sigma$ is the standard deviation) with a green and red line respectively.
The $\mu + 10 \sigma$ threshold is shown in blue (if this threshold lies within the range displayed in the panels). It is clearly visible that the probability distribution of $I$, tracked over time at a specific site, extends beyond the $\mu + 10 \sigma$ threshold, indicating the occurrence of super-extreme events. Here $4096$ spatial data points have been taken for a system of length $L = 150$.
%It has been done for $N=4096$ but the 
Qualitatively similar features are obtained for $2048$ spatial data points as well.
} %Here, we have taken random perturbations about the $(1,1)$ state. }
\label{prob_2}
\end{figure}

We now analyze the statistical properties of the variables $A$ and $I$, for the illustrative dynamical regimes presented in Fig.~\ref{kymo}. The left panels of Figs.~\ref{prob_1}-\ref{prob_2} display the probability distribution of the values of the variables $A$ (top) and $I$ (bottom), at  a specific spatial location, tracked over long times, after initial transience. We denote this quantities as $P_t(A)$ and $P_t(I)$.
%$T=5000,L=150$.Here,
%Left side represents the probability distribution of $A,I$ at $x=70$ and Right side represents the probability distribution of $A,I$ at $t=480$. Here, $N_x=4096,T=50000,L=150$(Ignoring the transients).Here, 
The right panels of these figures display the probability distribution of the values of the variables $A$ (top) and $I$ (bottom), tracked over space, at a specific time instant.  We denote this quantities as $P_s(A)$ and $P_s(I)$. The average value $\mu$ is shown with a black line in these plots, and the $\mu+\sigma$ and $\mu + 4\sigma$ are shown in green and red respectively, where $\sigma$ is the standard deviation. It is clear from these results that the probability distributions have a significant tail extending beyond the threshold for extreme event detection. This implies that extreme events occur in the system, for the concentrations of the two species, both in time and in space.

Interestingly, the number of extreme events that arise for the two chemical species, $A$ and $I$, are different. The slow inhibitor-like species $I$ exhibits a longer tail in the probability distribution, in comparison to that for the fast activator-like species $A$. This indicates an important feature: there are significant differences in the occurrence of extreme events in the two chemical species, with the species $I$ having a higher probability of producing extreme events. 

We also show an even higher threshold, marking a level that is $10 \sigma$ beyond the mean. Crossing this very high threshold indicates the presence of extreme events that may be as labelled super-extreme events \cite{xiong2025super,bonatto2017extreme,chabchoub2012super,durairaj2025super}, as such events are exceedingly far from the average state. Interestingly, it is evident from these figures, that super-extreme events arise in the slow chemical species, manifested by the probability distribution function extending considerably beyond the $10 \sigma$ threshold. In the solitonic regime, super-extreme events arise in the temporal evolution as well as in the spatial profile (see Fig.~\ref{prob_1}). In the regime of spatiotemporal chaos they arise in the temporal evolution at a site. 

\begin{figure}
    \centering
    \includegraphics[width=0.85\linewidth]{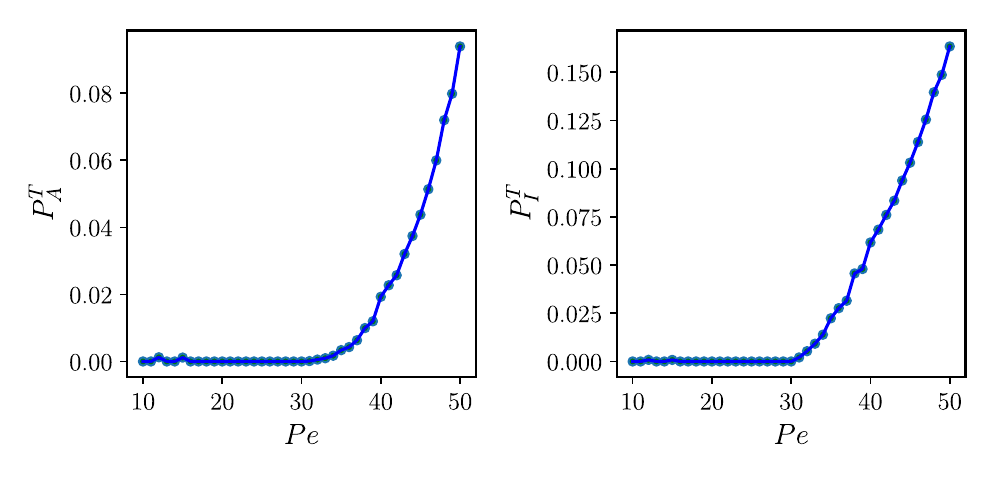}

\vspace{0.2cm}

    \includegraphics[width=0.85\linewidth]{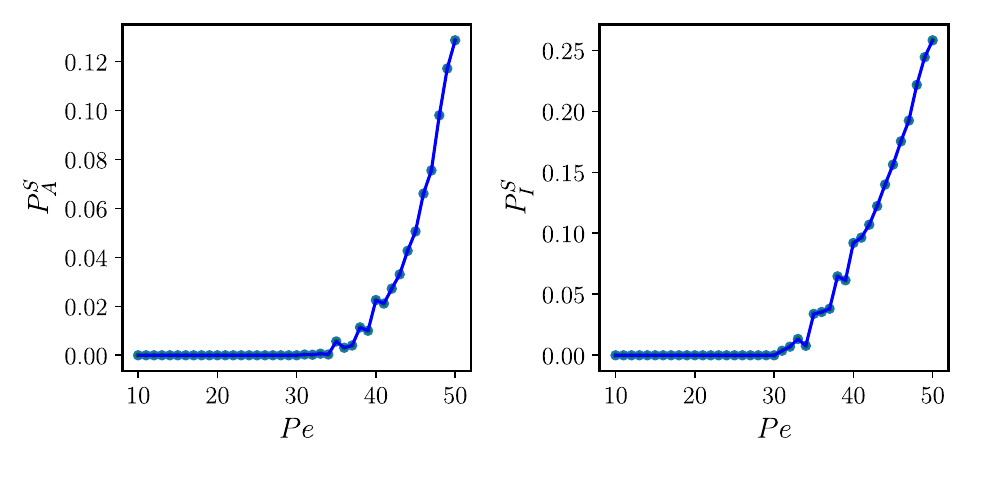}
    \caption{\raggedright Probability of temporal extreme events (top panels) and spatial extreme events (bottom panels) as a function of P\'eclet number Pe, for $A$ (left) and $I$ (right).}% Here $\beta=3$, $R=1$, $\alpha=0.1$ with random perturbations about $(1,1)$ state. }
    % Taken Time $T=280-300$.}
   \label{prob_EX}
\end{figure}

%\begin{figure*}[ht]
    %\centering
   % \includegraphics[width=0.8\linewidth,height=0.8\linewidth]{TemporalEEforP43b3R1withHist.pdf}
    %\caption{ Upper row represent the Time series and histogram of temporal extreme events of $A$  at $Pe=43,\beta=3,R=1,\alpha=0.1$ and Lower row represent the time series and histogram of $I$ for $T=50$ to $T=5000$(The black, green and red line correspond to the mean($\mu$),mean+standard deviation($\mu+\sigma)$,mean+4standard deviation($\mu+4\sigma$)}
    %\label{Pe_R}
%\end{figure*}

We also quantify the probability of observing extreme events, as the probability that a local maxima of the time series exceeds the prescribed threshold. The probability of the occurrence of temporal extreme events is denoted by $P^T_A$ and $P^T_I$, and the probability of the occurrence of spatial extreme events is denoted by $P^S_A$ and $P^S_I$, for the two chemical species $A$ and $I$ respectively. The variation of these probabilities with respect to P\'eclet number is shown in Fig.~\ref{prob_EX}. It is clear that there is a sudden onset of extreme events after a critical value of the P\'eclet number. This corresponds to the point in the bifurcation diagram (shown in Fig.~\ref{Bifurcationwithscaling}) where the system transitions to spatiotemporal chaos accompanied by an extremely large expansion in the size of the attractor. So one can infer that the P\'eclet number needs to be sufficiently large in order for the active fluid to exhibit extreme events. Increasing the P\'eclet number beyond the critical value,  leads to increasingly enhanced probabilities of encountering extreme events.

Note that we have also checked the generality of the trends by raising the threshold for extreme event detection to $\mu+6\sigma$, and we find that qualitatively similar behaviour is obtained. Further note that in the oscillatory regime there exist initial conditions that yield quasi-periodic behavior where the maximum amplitude exceeds threshold $H$. However, since those events are correlated, they do not qualify as extreme events, and are not counted in these measures.

\begin{figure}[htbp]
\centering

% ---------- Row 1 ----------
%\makebox[\textwidth][c]{%
%\includegraphics[width=0.22\textwidth]{IEIforAP48b3R1uptoT5000TEEseed42fixed.pdf}
\includegraphics[width=0.22\textwidth]{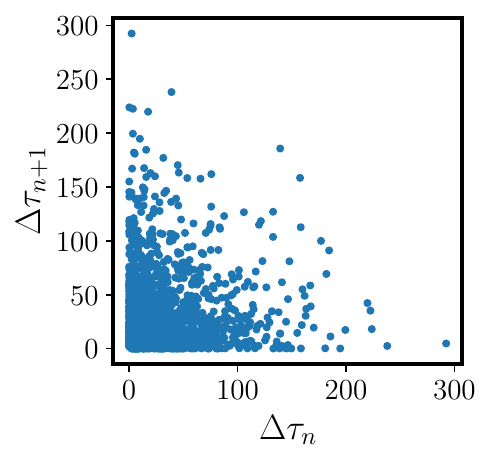}
\includegraphics[width=0.23\textwidth]{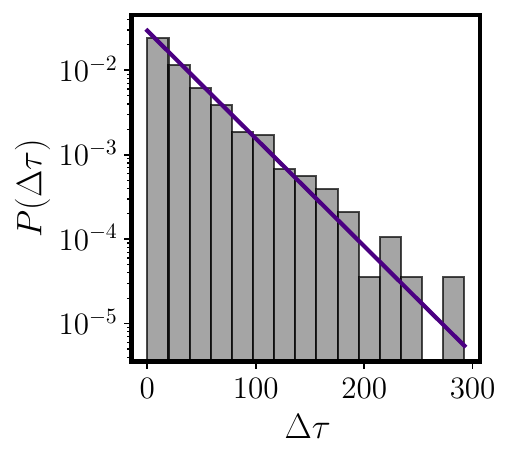}
%\includegraphics[width=0.22\textwidth]{P(I)vsIP48b3R1atx70uptoT5000.pdf}
%\includegraphics[width=0.22\textwidth]{P(I)vsIP48b3R1atx70uptoT5000spatialEE.pdf}
%}

\vspace{0.5cm}

% ---------- Row 2 ----------
%\makebox[\textwidth][c]{%
%\includegraphics[width=0.22\textwidth]{IEIforIP48b3R1uptoT5000TEEseed42fixed.pdf}
\includegraphics[width=0.22\textwidth]{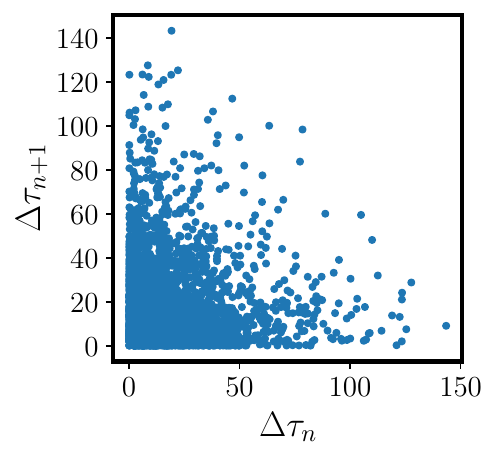}
\includegraphics[width=0.23\textwidth]{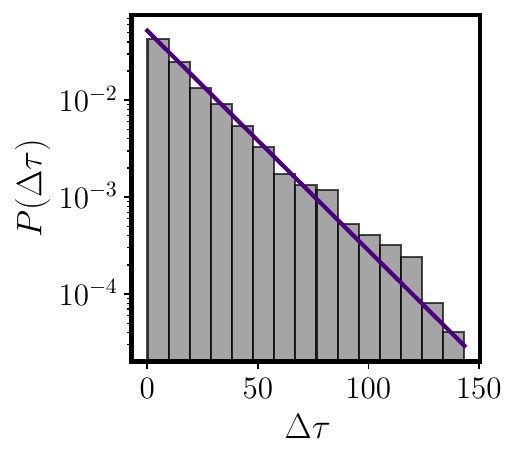}
%\includegraphics[width=0.215\textwidth]{AvstP48b3R1atx70uptoT5000.pdf}
%\includegraphics[width=0.225\textwidth]{AvsxP48b3R1att480uptoT500.pdf}
%}

\caption{\raggedright (Left panels) Return maps of inter-event intervals (IEI) of temporal extreme events for $A$ and $I$ respectively, and (Right panels) Probability distribution of inter-event intervals (IEI) of $A$ (top) and $I$ (bottom). Here $Pe=48$, $\beta=3$, $R=1$, $\alpha=0.1$. In the right panels, the indigo line depicts the distribution $P(\Delta \tau)=\lambda \exp(-\lambda\Delta\tau)$ in log-linear scale, where $\lambda$ represents the reciprocal of the average value of inter-event intervals. Here $\lambda_1=0.029$, $\lambda_2=0.052$, for $A$ and$I$ respectively.}%Here,we have taken random perturbations about the $(1,1)$ state.}
\label{IEI}
\end{figure}

In order to demonstrate that the extreme events are uncorrelated, we investigate the distributions and return maps the inter-event interval (IEI) denoted by $\Delta \tau$. The results for both variables $A$ and $I$ are displayed in Fig.~\ref{IEI}, for extreme events arising in the regime of spatiotemporal chaos. It is clearly evident from the return maps that the instances of extreme event occurrences exhibit no correlated patterns. Further, the distribution of the IEI follows a Poisson distribution, i.e $P (\Delta \tau)$ falls approximately as an exponential centered around zero.
%{\bf Bunching of Extreme events}:  

\begin{figure}
    \centering
    \includegraphics[width=\linewidth]{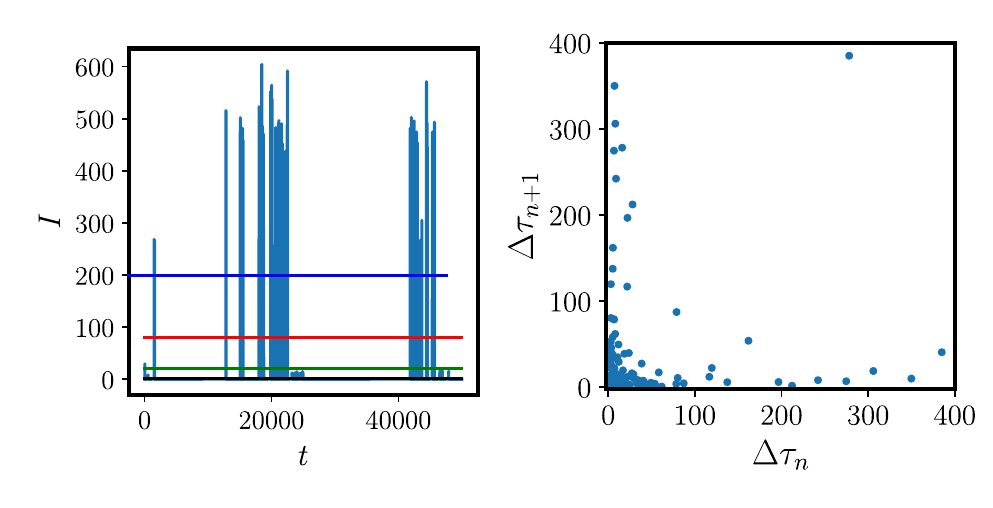}

\ \ \ \ \ \ \ \ \ \ \ \ \ \ \ \ \ \ \ \ \  (a) \hfill (b)   \ \ \ \ \ \ \ \ \ \ \ \ \ \ \ \  
     
    \caption{\raggedright (a) Time series and (b) Return map of inter-event intervals IEI, of the concentration of the slow-diffusing chemical species $I$. Here the system is in the regime of solitonic merging-emerging dynamics, with parameters $Pe=24$, $\beta=1.2$, $R=1$, $\alpha=0.1$. Notice that the bunching of extreme events is clearly visible in the time series. The $\mu+\sigma$, $\mu + 4\sigma$ and $\mu + 10 \sigma$ thresholds (where $\sigma$ is the standard deviation) are shown by green, red and blue lines respectively. It is clearly visible that the probability distribution of $I$, tracked over time at a specific site, extends beyond the $\mu + 10 \sigma$ threshold, indicating the occurrence of super-extreme events.}
    \label{Bunching5}
\end{figure}

Interestingly, in the regime of merging-emerging soliton-like dynamics 
%Bunching phenomena is clearly visible in spatio-temporal chaos and merging-emerging dynamics patterns. In the parameter regime where merging-emerging dynamics occurs, 
the extreme events in the concentration of the chemical species $I$ exhibits pronounced bunching in temporal extreme events. This manifests itself in the following manner: We observe no events for a long time, followed by a sudden burst of several spikes over a short interval of time, as evident through the time-series and return map of the inter-event interval displayed in Fig.~\ref{Bunching5}. In these figures bunching is reflected in the enhanced probability of very short inter-event intervals, resulting in a notable clustering around very small values in the return map of successive IEI, indicating that a short inter-event interval has a strong chance of being followed by another short inter-event interval. This leads to a bunching of extreme events occurring in a short span of time. The return map also shows that almost no long inter-event interval is followed by another long inter-event interval. This suggests that if there is a long gap between two extreme events, it is unlikley that the subsequent extreme event will occur after a long gap. In fact, the presence of points in the return map corresponding to a large $n^{th}$ IEI and a short ${(n+1)}^{th}$ IEI, indicates that the subsequent extreme event will likely emerge within a short interval. 
%Bunching is observed in the dynamics of both variables $A$ and $I$, though it is more prominent in the concentration of the slower species $I$, as evident in Fig.~\ref{Bunching2}.

%\begin{figure}
%    \centering
%    \includegraphics[width=\linewidth]{IEIforAandIP24b1_2R1uptoT50000TEEseed42fixedmerged.pdf}
%    \caption{\raggedright  Return maps of inter-event intervals (IEI) of temporal extreme events for $A$ and $I$ respectively. Here $Pe=24$, $\beta=1.2$, $R=1$, $\alpha=0.1$.}
%    \label{Bunching1}
%\end{figure}

%\begin{figure}
%    \centering
%    \includegraphics[width=\linewidth]{AvstandIvstforP24b1_25crorepointstransient50000pointsdseed42x103mergedbunching.pdf}
%    \caption{\raggedright Time series of concentrations (left) $A$ and (right) $I$. Here $Pe=24$, $\beta=1.2$, $R=1$, $\alpha=0.1$.}
%    \label{Bunching2}
%\end{figure}
%\begin{figure*}
    %\centering
   % \includegraphics[width=\linewidth]{P43b3R1IntereventandP(tau)forAandI.pdf}
   % \caption{(a),(c) represent the return maps of inter-event intervals of temporal extreme events for $A$ and $I$ respectively and (b),(d) represent the probability distribution of inter-event intervals of $A$ and $I$ for $Pe=43,\beta=3,R=1,\alpha=0.1$.}
    %\label{dispersion}
%\end{figure*}

\begin{figure}
    \centering
    \includegraphics[width=\linewidth]{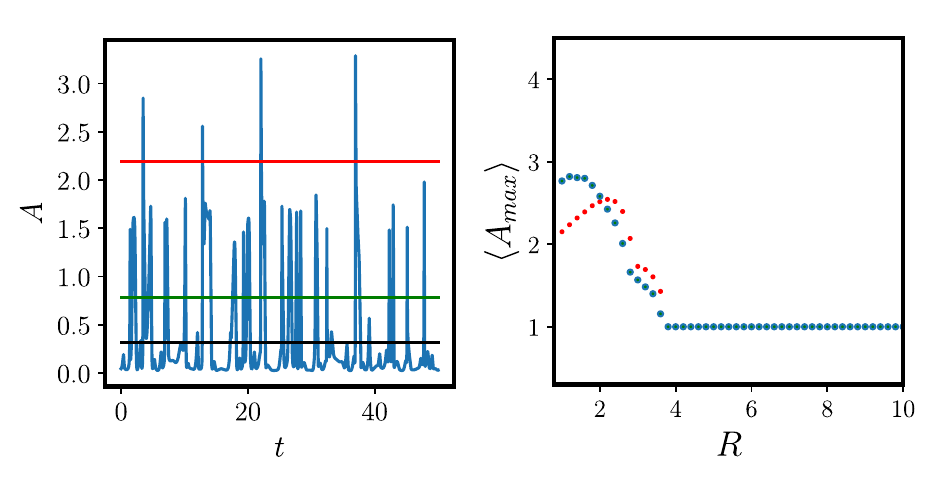}

 \ \ \ \ \ \ \ \ \ \ \ \ \ \ \ \ \ \ \ \ \  (a) \hfill (b)   \ \ \ \ \ \ \ \ \ \ \ \ \ \ \ \  
 
    \caption{\raggedright (a) Time series tracked over time at a specific site, for nonlinear parameter $R=1$, with $\mu$, $\mu+\sigma$ and the $\mu + 4\sigma$ threshold shown by black, green and red lines respectively. Here the system is in the regime of solitonic merging-emerging dynamics. (b) Variation of site averaged value of global maxima $\langle A_{max} \rangle$ of the concentration of the  chemical species $A$ (shown in blue), with respect to the nonlinear parameter $R$, and the threshold $H = \mu + 4 \sigma$ for determining extreme events (shown in red). The other parameters in (a-b) are: $Pe=24$, $\beta=1.2$, $\alpha=0.1$. The suppression of extreme events with increasing nonlinearity $R$ is evident, as $\langle A_{max} \rangle$ is significantly above the threshold $H$ only for low values of $R$.}
    \label{EE_withRpde}
\end{figure}

Lastly, we investigated the impact of nonlinearity on the emergence of extreme events. We find that increasing nonlinearity, as reflected by the parameter $R$, suppresses extreme events. This is evident through comparative study of the time series arising for varying values of $R$. It is also clear from the inspection of the site-averaged global maxima of the chemical species with respect to nonlinearity $R$, as compared to the threshold $H = \mu + 4 \sigma$ for determining extreme events (see Fig.~\ref{EE_withRpde}). The range of values of $R$ for which this global maxima lies significantly above the threshold $H$ marks the window where discernible extreme events occur. Fig.~\ref{EE_withRpde} also shows a representative time series for the system with $R=1$, tracked at a site, displaying uncorrelated instances of the concentration of $A$ exceeding the $4 \sigma$ threshold.

\section{Analysis of a Reduced-Order Model }

We now analyze a reduced-order model, motivated by the Galerkin model of the PDEs describing the system. The set of ODEs we consider have been obtained by applying a Galerkin projection~\cite{fletcher1984computational} to approximate the temporal evolution of the dominant modes, with additional stabilizing nonlinear terms. The ODE system we study below shares qualitative similarity with the dynamics of the original system.

The evolution equations for the amplitudes \( A_i, I_i \), using $f(A,I) =
(1+\beta)\frac{A}{A_S} + (1-\beta)\frac{I}{I_S}$, are:

\begin{align}
\frac{d X_i}{dt} &= \sum_{jklm} c^i_{jlmn} X_j^m X_l^n \nonumber \\
%\frac{dI_i}{dt} &= \sum_{jklm} d^i_{jlmn} I_j^m A_l^n  
\label{ROM}
\end{align}
%\begin{align}
%\frac{dA_i}{dt} &= \sum_{jklm} c^i_{jlmn} A_j^m I_l^n \nonumber \\
%\frac{dI_i}{dt} &= \sum_{jklm} d^i_{jlmn} I_j^m A_l^n  
%\end{align}
\noindent
Here $i,j,l=1,6$ are the indices for the variables, with $X_1=A_1$, $X_2=A_2$, $X_3=A_3$,$X_4=I_1$,$X_5=I_2$,$X_6=I_6$. The constant coefficients $c^i_{jlmn}$, where $m,n=0, 1$, are functions of the parameters $\mathrm{Pe}$, $k$, $\alpha$, $\beta$, $R$, $A_S$ and $I_S$. Coefficients where either index $m$ or $n$ is zero, imply linear terms. A linear term reflects self-coupling interaction with the same component if $i=j, m \ne 0, n=0$, and cross-coupling between dissimilar components and variables, otherwise. Non-zero coefficients $c^i_{jlmn},  \ne 0$, where both $m \ne 0$ and $n \ne 0$ indicate the presence of nonlinear terms. (See Appendix for details.)

%\begin{align}
%\frac{dA_1}{dt} &= c_1 A_1  + {c_2 A_1A_2} + {c_3 A_2A_3} + {c_4 I_1}+ {c_5 A_3I_2}  \nonumber \\
% &  + {c_6 A_1 I_2} + {c_7 A_2 I_3} + {c_8 A_2 I_1}
%\\[8pt]
%\frac{dA_2}{dt} &= c_9 A_2+ c_{10}A_1^2 + c_{11}A_1A_3 + {c_{12}A_1I_1} \nonumber \\
%& + {c_{13}A_3I_1}+{c_{14}A_1I_3} + {c_{15}I_2}, \nonumber
%\\[8pt]
%\frac{dA_3}{dt} &= c_{16}A_3+ {c_{17}A_1A_2}+ {c_{18}A_2I_1} \nonumber \\
%5& + {c_{19}A_1I_2} + {c_{20}I_3} .
%\end{align}

%\begin{align}
%\frac{dI_1}{dt} &= d_1A_1+ {d_2I_1}+ {d_3I_1A_2}+ {d_4I_1I_2}+ {d_5I_2A_3} \nonumber \\
%& + d_6 I_2 I_3 + d_7 I_2 A_1 + d_8 I_3 A_2 \nonumber
%\\[8pt]
%\frac{dI_2}{dt} &= d_9 A_2+ d_{10}I_2 + d_{11}I_1 A_1 +{d_{12}I_1^2} \nonumber \\
%& + d_{13} I_3 A_1 + d_{14}I_3 I_1 + d_{15}I_1 A_3  \nonumber
%\\[8pt]
%\frac{dI_3}{dt} &= d_{16} A_3 + {d_{17} I_3}+ {d_{18} I_2 A_1} \nonumber \\
%&+ d_{19} I_2 I_1 + d_{20} I_1 A_2 .
%\end{align}

%where $c_i$, $i=1, \dots 20$ and $d_i$, $i=1, \dots 20$ are constants (see Appendix for details).

\begin{figure}
    \centering
    \includegraphics[width=0.7\linewidth]{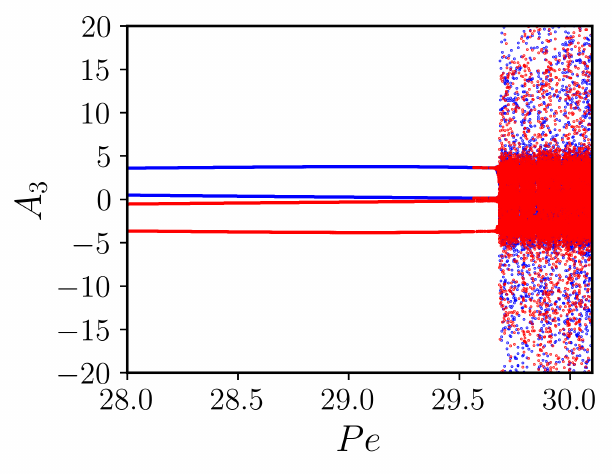}
    \caption{\raggedright Bifurcation diagram displaying the local maxima and minima of variable $A_3$ with respect to varying P\'eclet number $Pe$, arising in the system given by Eqns.~5-7, for different sets of initial conditions. Two coexisting limit-cycle attractors (colored in red and blue) are clearly visible. These eventually collide, and give rise to a single wide chaotic attractor around $Pe=29.8$.  Here $\beta=2$, $R=1$, $\alpha=0.1$, $A_S=8,I_S=8$, $L=30$. 
    %The(Blue dots represent local maxima and red dots represent local minima)(Taken $1000$ data points of $Pe$)(Taken Initial conditions as $(1.0017,1.0017,1.0017,1.0017,1.0017,1.0017)$
    }
    \label{bif_multistability}
\end{figure}

\begin{figure}
    \centering
    \includegraphics[width=\linewidth]{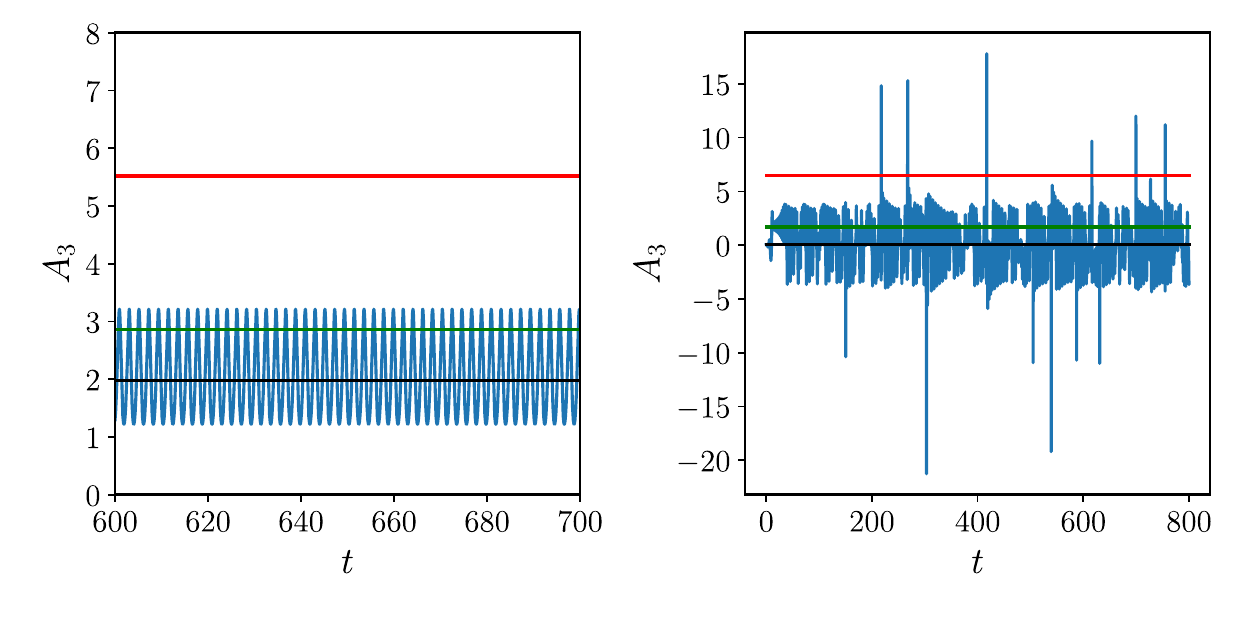}
    \includegraphics[width=\linewidth]{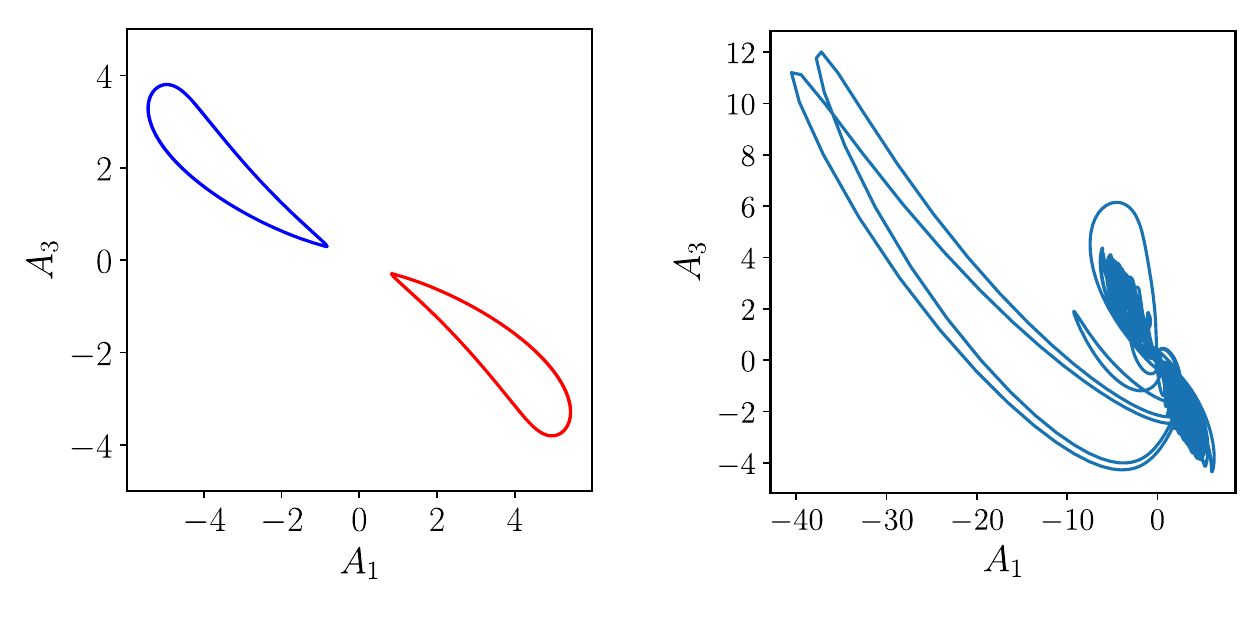}
 
    \caption{\raggedright Dynamics of the reduced-order system given by Eqns.~5-7: (left) regular oscillations for $Pe=29$, and (right) chaotic oscillations for $Pe=31$, with $\beta=2$, $R=1$, $\alpha=0.1$. The top panels show the times series of variable $A_3$, with the average value $\mu$ of the variable shown with a black line, and $\mu+\sigma$ and $\mu + 4\sigma$ (where $\sigma$ is the standard deviation) with a green and red line respectively. The lower panels show the attractors in $A_1-A_3$ space, clearly showing the differences in attractor size. Red and blue attractors in the left side are obtained for two distinct initial conditions.}
    \label{ROM_timeseries}
\end{figure}

%\subsection{Multistability}
In this work, we consider $A_S=I_S=8$ , $L=30$ and $k = \frac{2\pi}{L}$. We display the bifurcation diagram of the system above, over a range of P\'eclet numbers, for different sets of initial conditions in Fig.~\ref{bif_multistability}. It is clearly evident from the bifurcation diagram that there is multistability in this system, and two coexisting limit-cycle attractors (colored in red and blue) are clearly visible. Representative time series and corresponding phase space attractors of the two qualitatively different behavioral regimes are presented in Fig.~\ref{ROM_timeseries}. The basin of attraction of these limit cycles are separated by an unstable manifold. As the P\'eclet number increases, these limit cycles approach each other and eventually collide at a saddle point. This gives rise to a single wide chaotic attractor around $Pe \sim 29.8$. 

\begin{figure}
    \centering
    \includegraphics[width=0.6\linewidth]{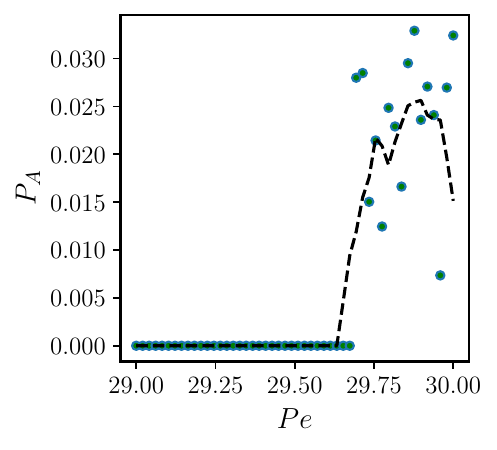}
    \caption{\raggedright Probability of temporal extreme events in the variable $A_3$, with respect to P\'eclet number $Pe$, for the reduced-order model. Here, $A_S=I_S=8$, $\beta=2$, $R=1$, $\alpha=0.1$, $L=30$. Here the probability is estimated through time averages (after transience) and ensemble average over $20$ random initial conditions. The black dotted line represents the $6$-point running average. 
    %Taken time upto $T=1800$(ignoring the transients).
    }
    \label{probofTEEROM}
\end{figure}

\begin{figure}
    \centering
    \includegraphics[width=\linewidth]{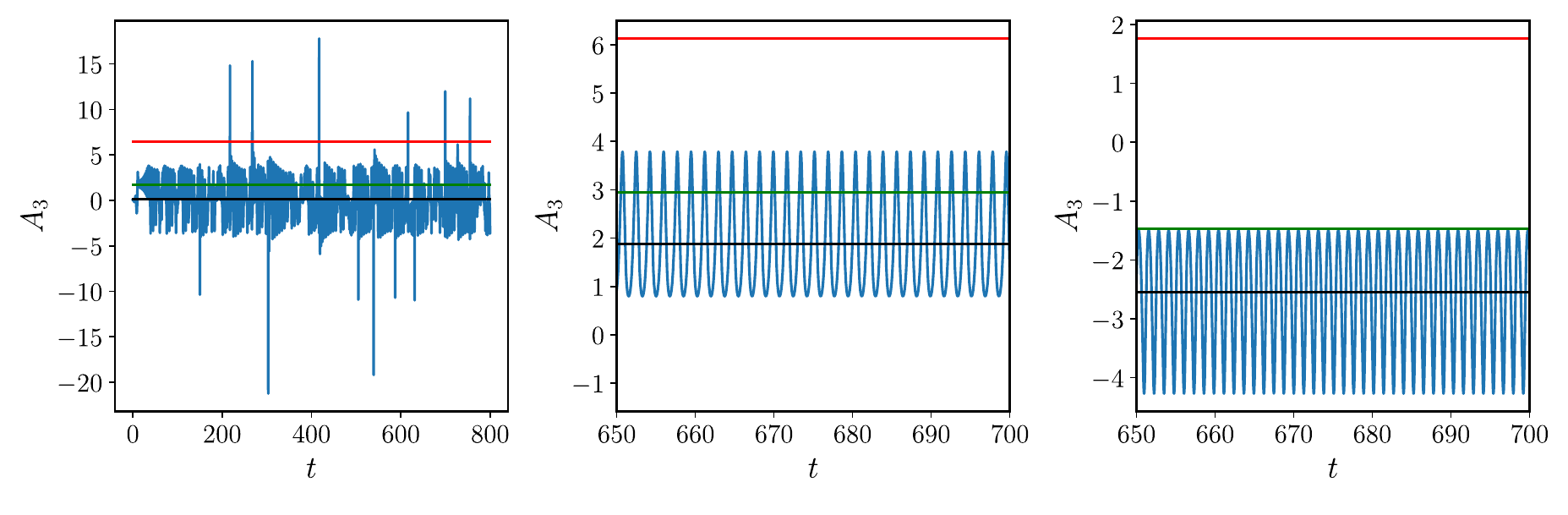}
    \caption{\raggedright Suppression of extreme events with increasing $R$. Here $Pe=31$,$\beta=2$,$\alpha=0.1$,$L=30$,$A_S=I_S=8$, and left to right:  $R=1$, $1.25$, $1.5$. The average value $\mu$ is shown with a black line, and $\mu + \sigma$ and $\mu + 4\sigma$ (where $\sigma$ is the standard deviation) with a green and red line respectively.}
    \label{EE_ROM}
\end{figure}

So this crisis leads to a sudden attractor expansion clearly evident in the bifurcation diagram, and is the underlying mechanism for the emergence of extreme events in this model. We quantify this trend again by estimating the probability of obtaining extreme events, with increasing P\'eclet number. The results are given in Fig.~\ref{probofTEEROM}, from where we can distinctly see that extreme events appear suddenly after the merger of the coexisting limit cycles and the subsequent attractor expansion. While increasing P\'eclet number aids extreme events, increasing the scaled nonlinearity parameter $R$ suppresses extreme events. We show illustrative examples of this suppression with increasing nonlinearity in Fig.~\ref{EE_ROM}.

\begin{figure}
    \centering
    \includegraphics[width=\linewidth]{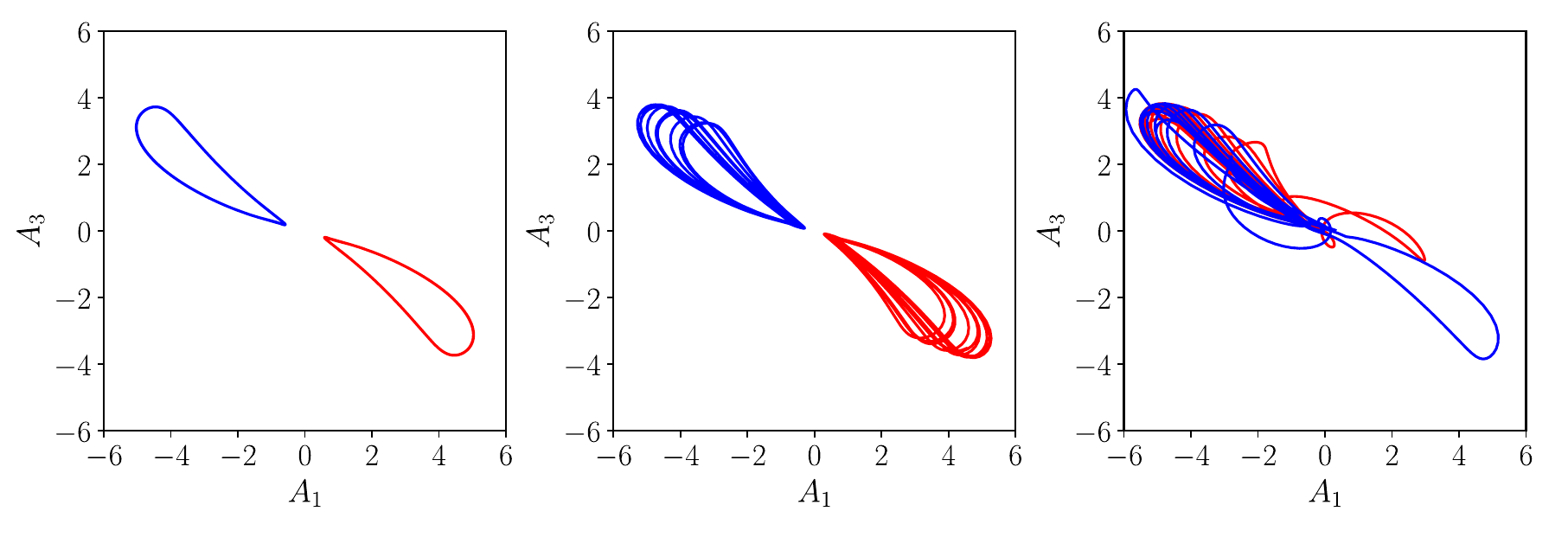}
    \caption{\raggedright Merging of two limit-cycle attractors with changing $Pe$. Here $R=1$, $\beta=2$, $\alpha=0.1$, $L=30$, $A_S=I_S=8$, and left to right:  $Pe=29.5$, $29.67$, $29.7$.}
    \label{EE_attractormerging}
\end{figure}
%\begin{figure}
%    \centering
%    \includegraphics[width=\linewidth]{A3vsA1forP30ASIS8b2andP25AsIS8ROMmergednew.pdf}
%    \caption{Left image represents the limit cycle corresponds to $Pe=25,\beta=2,R=1,L=30,A_s=I_s=8,\alpha=0.1$ and Right image represents the chaotic attractor with large deviations for $Pe=31,\beta=2,R=1,L=30,A_s=I_s=8,\alpha=0.1$ }
%    \label{attractors_ROM}
%\end{figure}

Extreme events in low-dimensional dynamical systems can arise through a variety of well-established routes, including crisis-induced intermittency, Pomeau–Manneville intermittency, quasiperiodicity-induced intermittency, and noise-induced intermittency. These mechanisms have been extensively discussed in \cite{chowdhury2022extreme}. In the present system, the emergence of extreme events is primarily associated with an attractor-merging crisis. This mechanism is clearly identifiable in both the bifurcation diagrams and the phase-space portraits. As the P\'eclet number ($Pe$) increases, the system reveals the progressive approach and eventual collision of two coexisting attractors. This global event leads to the merging of the two distinct limit cycle attractors into one larger chaotic attractor, and the formation of unstable manifolds. When system trajectories approach these unstable manifolds in phase space, they are rapidly repelled, resulting in large excursions away from the typical attractor region. These excursions manifest as extreme events and correspond to a sudden expansion of the chaotic attractor. Subsequently, due to the presence of attracting directions, trajectories are re-injected into the nominal attractor region \cite{sudharsan2025extreme1,karnatak2014route}.
%, restoring the system orbits to its typical dynamical region.
%A key geometric feature associated with these extreme events is the presence of a narrow channel-like structure in phase space. Extreme trajectories consistently pass through this confined region during both their departure from, and return to, the nominal attractor. Notably, different extreme-event trajectories follow similar pathways through this channel, indicating a common dynamical mechanism. Such narrow escape channels are a characteristic feature of systems exhibiting extreme events and have been reported in earlier studies \cite{sudharsan2025extreme1,karnatak2014route}. %Also, In our original system, the velocity field is nonlocal in nature, clearly visible from the structure of the equation. It is reported in many studies that nonlocal coupling and nonlinearity are two important ingredients that are responsible for the emergence of rogue waves/extreme events\cite{onorato2013rogue,horikis2017rogue}. Extreme events such as Rogue waves in optics and fluids are often associated with the merging dynamics of coherent structures\cite{selmi2016spatiotemporal}.

\section{Conclusions}

 In this work we explored the existence of extreme events in an active fluid, involving two distinct chemical species that regulate active stress. One species is slow diffusing and the other is fast diffusing, and the growth of the fast-diffusing species is modelled using a nonlinear logistic term. %We  investigated the occurrences of extreme event under varying P\'eclet number and strength of the nonlinear growth term, and 
 We find distinct evidence of rare and uncorrelated extreme concentration build-ups, in regimes of spatiotemporal chaos and merging-emerging soliton-like dynamics. We characterized these events through the examination of the time-series, bifurcation diagrams and the probability distribution functions of the concentration of the two species, and clearly demonstrated the emergence of extreme events in the temporal evolution of the concentrations, as well as in the spatial profile of the system. 
 
 Further, we systematically explored the dependence of extreme event occurrences on the two important parameters in the system: the scaled non-linear growth parameter $R$, which reflects the strength of the nonlinear growth term, and the P\'eclet number $Pe$, which reflects the ratio of the advective transport to diffusive transport in fluid flow. We find that the P\'eclet number and the strength of the nonlinear growth have contrasting impact on extreme events. The probability of extreme events increases after a critical P\'eclet number, i.e. increasing P\'eclet number aids the generation of extreme events. On the other hand, increasing nonlinearity suppresses extreme events, quenching them completely for sufficiently strong nonlinearity.
 
Interestingly, we also find evidence of pronounced bunching of extreme events, in the slow chemical species in the soliton-like regime. This is  manifested as the occurrence of extreme events in quick succession following a long period of quiescence, and this feature is clearly revealed through examination of the inter-event interval distributions and return maps. Further, the extreme events in this region can be considered as super-extreme, as they are more than $10 \sigma$ away from the mean.

Note that the extreme events in the solitonic regime of the active fluid system are reminiscent of rogue waves observed in optical and oceanic systems, as they are linked to the interaction and merging dynamics of coherent structures \cite{selmi2016spatiotemporal}. Also note that the velocity field in our system is intrinsically nonlocal in nature, as is evident from the structure of the governing equations, and the interplay between nonlocal coupling and nonlinearity has often constituted a key mechanism underlying the emergence of rogue waves \cite{onorato2013rogue,horikis2017rogue}. 

Lastly, in order to gain further understanding of the extreme events, we investigate a modified mode-truncated reduced order model of the system, and demonstrate that it also exhibits temporal extreme events. The emergence of these events is correlated with the sudden expansion in attractor size, as evident from the bifurcation diagram and phase plots of the attractors, offering insight on the mechanism underlying the emergent events. 

In summary, our results here clearly demonstrate the existence of distinct temporal and spatial extreme events in an active fluid system. Such events are manifested as extreme concentration buildups at a site at rare instances in its temporal evolution, or at a few sites in the spatial profile of the system at an instant of time. The presence of such events in an active fluid system can be anticipated to have a potentially crucial impact on biological phenomena where active transport plays an important role.

\bigskip
\bigskip

\section{Appendix}

%Considering all possible combinations of $jlmn$,  where $j,l$ can take 6 values and $m,n$ can take 2 values ($0$ or $1$), there are a very large number of possible coefficients $c^i_{jlmn}$. However, we find that only a small subset of these are non-zero.

In this Appendix we give the full set of dynamical equations in our reduced order model analyzed in Section V. Specifically, in this work we have considered $\beta=2$, and $A_S=I_S$. For this case, the parameters may be regrouped, such that all the non-zero coefficients can be compactly expressed in terms of just three quantities:

$$\epsilon_1=\frac{\mathrm{Pe}\,k^2}{(1+k^2)A_S}, \ \epsilon_2=\frac{4\,\mathrm{Pe}\,k^2}{(1+4k^2)A_S}, \ \epsilon_3=\frac{9\,\mathrm{Pe}\,k^2}{(1+9k^2)A_S}$$ 

\medskip

Using the quantities given above, the dynamical equations given in Eqn.~\ref{ROM} become:

\begin{widetext}
\begin{align*}
\frac{dA_1}{dt} &= (3 \epsilon_1 - k^2 - R) A_1  -(\frac{9}{2}\epsilon_1 + \frac{9}{4}\epsilon_2 + R) \ A_1A_2 - ( \frac{15}{4} \epsilon_2 +  \frac{5}{2} \epsilon_3 + R) \ A_2A_3 - \epsilon_1 I_1+ \frac{5}{4} \epsilon_2 \ A_3 I_2 \nonumber + \frac{3}{4} \epsilon_2 A_1 I_2 \\
& \ \ \ \ \ \ \ \ \ + \frac{5}{6} \epsilon_3 A_2 I_3 + \frac{3}{2} \epsilon_1 A_2 I_1 \nonumber \\
\\[8pt]
\frac{dA_2}{dt} &= (3 \epsilon_2-4k^2-R) A_2+ (3 \epsilon_1-\frac{R}{2}) A_1^2  -(3 \epsilon_1 - \epsilon_3 + R ) A_1 A_3 - \epsilon_1 A_1 I_1
+ \epsilon_1 A_3 I_1 -\frac{\epsilon_3}{3} A_1 I_3 -\epsilon_2 I_2, \nonumber\\
\\[8pt]
\frac{dA_3}{dt} &= (3 \epsilon_3-9k^2-R ) A_3+ \frac{9}{2} (\epsilon_1 + \frac{\epsilon_2}{2}- \frac{2}{9}R) A_1A_2 -\frac{3}{2} \epsilon_1 A_2 I_1 -\frac{3}{4} \epsilon_2 A_1 I_2 -\epsilon_3 I_3 \\[12pt]
\frac{dI_1}{dt} &=  3 \epsilon_1 A_1-(\epsilon_1 + k^2 \alpha) I_1 -\frac{9}{4} \epsilon_2 I_1 A_2+ (\frac{3}{2} \epsilon_1 + \frac{3}{4} \epsilon_2) I_1 I_2 - \frac{5}{2} \epsilon_3 I_2 A_3 + \frac{5}{2} (\frac{\epsilon_2}{2}+\frac{\epsilon_3}{3}) I_2 I_3 - \frac{9}{2} \epsilon_1 I_2 A_1 - \frac{15}{4} \epsilon_2 I_3 A_2 \nonumber \\
\\[8pt]
\frac{dI_2}{dt} &= 3 \epsilon_2 A_2 -(\epsilon_2+4k^2\alpha)I_2 + 3 \epsilon_1 I_1 A_1 -\epsilon_1 I_1^2  -3 \epsilon_1 I_3 A_1 + (\epsilon_1 -\frac{\epsilon_3}{3}) I_3 I_1 + \epsilon_3 I_1 A_3 \nonumber\\
\\[8pt]
\frac{dI_3}{dt} &= 3 \epsilon_3 A_3 -(\epsilon_3 + 9k^2\alpha) I_3+ \frac{9}{2} \epsilon_1 I_2 A_1 - (\frac{3}{2} \epsilon_1 + \frac{3}{4} \epsilon_2) I_2 I_1 + \frac{9}{4} \epsilon_2 I_1 A_2 
\end{align*}
\end{widetext}

Section V analyzes the system described by the above  set of equations.

\bibliography{Extreme2}% Produces the bibliography via BibTeX.
\bibliographystyle{apsrev4-2}
\end{document}